\documentstyle[12pt]{article}
\begin{document}
\pagestyle{plain}
\title{Special light trajectories in optical medium}
\author{Miroslav Pardy\\
Department of Physical Electronics \\
Masaryk University \\
Kotl\'{a}\v{r}sk\'{a} 2, 611 37 Brno, Czech Republic\\
e-mail:pamir@physics.muni.cz}
\date{\today}
\maketitle
\vspace{50mm}

\large

\begin{abstract}
The Fermat principle is used to define trajectories in nonhomogenous optical media. The Poincar\'e model of the Loba\v cevskii geometry is derived. The index of refraction is determined for the light confined in the circular trajectory in the optical medium.  
\end{abstract}

\vspace{1cm}


\section{Introduction}

Trajectories of elementary particles are the basic ingredients of physics of elementary particles and cosmical rays. No elementary particle can exist without its trajectory. While in  particle physics the trajectories of particles are determined by their parameters as mass, charge, spin, velocity and by the influence of the magnetic and electric fields on its motion, the trajectories of light in geometrical optics are determined by the index of refraction of the optical media. 

Geometrical optics considers the propagation of waves of light, as the propagation of rays, completely divorced from their wave properties. 
In other words, geometrical optics corresponds to the limiting case of small wavelength, $\lambda \rightarrow 0$. 
 
The fundamental equations of geometrical optics for the direction of the rays are derived for instance by Landau et al. (1988, 1982) for  any quantity  $f$ describing the field of 
the wave (any component of ${\bf E}$ or ${\bf H})$. For a plane monochromatic wave, we have for $f$ the form 

$$f=ae^{i({\bf k.r}-\omega t +\alpha)} =ae^{i(-k_{r}x^{r} + \alpha)},\eqno(1)$$ 
(we omit the Re as the real part of all expressions). 
We write the expression for the field in the form 

$$ f = ae^{i\psi}.\eqno(2)$$ 
 
In case the wave is not plane, but geometrical optics is applicable, the amplitude $a$ is, 
generally speaking, a function of the coordinates and time, and the phase $\psi $ which is called 
the eikonal, does not have a simple form, as in (1). It is essential, however, that  $\psi$ be a 
large quantity. This is clear immediately from the fact that eometrical optics corresponds to the limit $\lambda \rightarrow 0$. 
 
Over small space regions and time intervals the eikonal $\psi$ can be expanded in series. The time derivative 
of $\psi$ gives the frequency of the wave: 

$$\frac{\partial \psi}{\partial t} = -\omega \eqno(3)$$  
and the space derivatives give the wave vector 

$${\bf grad}\;\psi = {\bf k}\eqno(4)$$ 
and consequently the direction of the ray through any point in space. 
For a steady monochromatic wave, the frequency is a constant and the time dependence 
of the eikonal is given by a term $-\omega t$. We then introduce a new function $\psi_{1}$ (also called the eikonal), such that 

$$\psi = -\omega t + (\omega/c) \psi_{1}(x,y,z).\eqno(5)$$. 

Then $\psi_{1}$ is a function of the coordinates only, and its gradient is 

$${\bf grad}\;\psi_{1} = {\bf n},\eqno(6)$$  
where {\bf n} is a vector such that 

$${\bf k} = \omega {\bf n}.\eqno(7)$$ 
 
The magnitude of ${\bf n}$ is equal to the refractive index $n$ of the medium. Hence the 
equation for the eikonal in ray propagation in a medium with refractive index $n(x,y,z)$ ($a$ given function of the coordinates) is 

$$|{\bf grad}\;\psi_{1}|^{2} =  \left(\frac{\partial\psi_{1}}{\partial x}\right)^{2} +
\left(\frac{\partial\psi_{1}}{\partial y}\right)^{2} +
 \left(\frac{\partial\psi_{1}}{\partial z}\right)^{2} =   n^{2}.     \eqno(8)$$  

The equation of ray propagation in a steady state can also be derived from Fermat's principle, according to which the integral 
 
$$\psi_{1} = \int_{A}^{B}{\bf n}.d{\bf l} = \int_{A}^{B}ndl\eqno(9)$$  
along the path of the ray between two given points $A$ and $B$ has a value less than for any other path between $A$ and $B$. 

Equating to zero the variation of this integral, we have 
 
$$\delta\psi_{1} = \int_{A}^{B}({\delta n}.d{l} +n\delta dl) = 0.\eqno(10)$$  

Let  $\delta {\bf r}$ be a displacement of the ray path under the variation. Then $\delta n = \delta{\bf r}.{\bf grad}\; n$, $\delta dl = {\bf 1}.d\delta{\bf r}$, where ${\bf 1}$ is a unit vector tangential to the ray. Substituting in $\delta \psi_{1}$ and 
integrating by parts in the second term (using the fact that $\delta {\bf r} = 0$ at $A$ and $B$), we have 

$$\delta \psi_{1} = \int_{A}^{B}\delta{\bf r} .{\bf grad}\;{n}\;d{l} +
\int_{A}^{B}n{\bf l}.d\delta {\bf r} = \eqno(11)$$ 

$$\int_{A}^{B}\left({\bf grad}\; n - \frac{d(n{\bf l})}{dl}\right).\delta {\bf r}dl. \eqno(12)$$ 

Hence 

$${\bf grad}\; n.d{l} = \frac{d(n{\bf l})}{dl}.\eqno(13)$$ 

Expanding the derivative and putting $dn/dl = {\bf 1}. {\bf grad}\; n$, we obtain
 
$$\frac{d{\bf 1}}{dl} = \frac{1}{n}[{\bf grad}\;n - 
{\bf 1}({\bf l} . {\bf grad}\;n)].\eqno(14)$$

This is the equation giving the form of the rays.
 
We know from differential geometry that the derivative $d{\bf 1}/dl$ along the ray is equal to 
${\bf N}/R$, where ${\bf N}$ is the unit vector along the principal normal and R the radius of curvature. 
Taking the scalar product of both sides of (14) with ${\bf N}$, and using the fact that ${\bf N}$ and ${\bf 1}$ are 
Perpendicular, we have
 
$$\frac{1}{R} = {\bf N}.\frac{{\bf grad}\;n}{n}. \eqno(15)$$

The rays are therefore bent in the direction of increasing refractive index. 

The velocity of propagation of rays in geometrical optics is in the direction of $\bf 1$ and is given by the derivative 

$${\bf u} = \partial\omega/\partial{\bf k}.\eqno(16)$$ 

This is also called the group velocity, the ratio $\omega/k$ being called the phase velocity. However, 
the latter is not the velocity of physical propagation of any quantity.
 
\section{Light trajectory from the Fermat principle}

The point $A$ and $B$ can be  joined by the infinite number of lines passing from $A$ to $B$. There is the shortest distance between $A$ and $B$ forming  the segment of the straight line passing through $A$ and $B$ (Hilbert, 1902).  The shortest distance between point $A$ and $B$ can be physically realized by the flexible but non-elastic fibre.  
 The segment $AB$ of a straight line can be prolongated in the direction $AB$, or $BA$ in order to generate the straight line.
The shortest trajectory $y = y(x)$ between two point $A(x_{1}, y_{1})$ and $B(x_{2}, y_{2})$ in the Euclidean plane x-y is $y = ax + b$, $a, b$ being some constants which  can be find as the solution of the Bernoulli izoperimetric problem of the variational calculus with functional

$$F = \int_{x_{1}}^{x_{2}}[1 + y'^{2}]^{1/2}dx ; \quad y' = \frac{dy}{dx}\eqno(17)$$
after insertion of it into the Euler-Lagrange equation

$$ \frac{\partial F}{\partial y} - \frac{d}{dx}\left(\frac{\partial F}{\partial y'}\right) = 0\eqno(18)$$
and after its solution (Lavrentjev et al., 1950).

 The geodetic line $y = y(x), z = z(x)$ from
point $A(x_{1}, y_{1}, z_{1})$ to point $B(x_{2}, y_{2}, z_{2})$ on the surface $\varphi(x, y, z) = 0$ is the solution of the izoperimetric problem with the functional

$$F = \int_{x_{1}}^{x_{2}}\left\{[1 + y'^{2} + z'^{2}]^{1/2} - \lambda 
\varphi(x, y, z)\right\}dx ; \quad y' = \frac{dy}{dx}, z' = \frac{dz}{dx},\eqno(19)$$
where $\lambda$ is the Lagrange multiplicator.

The  Fermat optical theorem states that the  trajectory of light from point $A$ to $B$ in the optical medium is of the shortest optical length. At the same
time the trajectory of the optical ray from point $A$ to point $B$ with reflection on the mirror  in point $C$ is of the shortest optical length in optical medium.

The trajectory of light passing from $A(x_{1}, y_{1})$ to $B(x_{2},
y_{2})$ is determined  by the Fermat principle, which states that the time from $A(x_{1}, y_{1})$ to $B(x_{2}, y_{2})$ is the result of the minimization of the functional 

$$T(y,y') = \int_{x_{1}}^{x_{2}}\frac{ds}{v(y)} = \int_{x_{1}}^{x_{2}}\frac{\sqrt{1 + y'^{2}}}{v(y)}dx, \eqno(20)$$
where $v(y)$ is the velocity of light.

The functional $T(y,y')$ is the solution of  the Euler-Lagrange equations (18).

If $v = Ay$, the solution of eq. (18) is in the form of circles forming the Poincar\'e model of the  Loba\v cevskii geometry:

$$(x - C)^{2} + y^{2} = r^{2}.\eqno(21)$$

The last equation can be also derived for velocity $v = Ax$. Then the functional (20) is of the form 

$$T(y,y') = \int_{x_{1}}^{x_{2}}\frac{ds}{v(x)} =
\int_{x_{1}}^{x_{2}}\frac{\sqrt{1 + y'^{2}}}{v(x)}dx, \eqno(22)$$

We get from the Euler equation with $y' = dy/dx$

$$T_{y} - \frac{d}{dx}T_{y'} = 0 \eqno(23)$$
we get $T_{y'} = const = C_{1}$, or 

$$ \frac{y'}{x\sqrt{1 + y'^{2}}} = C_{1}.\eqno(24)$$

We get with the substitution $t = \arctan y' $,

$$ x = \frac{y'}{C_{1}\sqrt{1 + y'^{2}}} = \frac{1}{C_{1}}\sin t =
C_{2}\sin t.\eqno(25)$$ 

It follows from $y' = \tan t$ that $dy = C_{2}\sin t dt$. Or

$$y = - C_{2}\cos t  + C_{3}.\eqno(26)$$
Or,

$$x = C_{2}\sin t, \quad y  - C_{3} = - C_{2}\cos t. \eqno(27)$$

If we  eliminate the variable $t$ in the last equation, we get 

$$ x ^{2} + (y - C_{3})^{2} = C_{2}^{2} = r^{2},\eqno(28)$$
which is a circle with the center on the y-axis. 
 
\section{Trajectory of light in the stratified medium}

Let us consider the stratified 2D medium in the plane $x-y$, where every layer is parallel with the x-axis. Then, $dx \rightarrow (\Delta x)_{i}, dy \rightarrow H/n$,  in (20) where i = 1, 2, 3, ..., and $H$ is the height of the stratified medium. In this case the velocity is $v_{i} = v(y_{0} + iH/n)$. Then, instead of functional $T(y,y')$ we write the following sum:

$$T = \sum_{i=0}^{\infty} \frac{\sqrt{((\Delta x)_{i})^{2} + (\Delta y)^{2}}}{v_{i}} = \sum_{i=0}^{\infty} \frac{\sqrt{(a_{i+1} - a_{i})^{2} + (H/n)^{2}}}{v_{i}}.\eqno(29)$$

The last sum has the stationary value if and only if $\partial T/\partial a_{i} = 0$. Or,

$$\partial T/\partial a_{i} = $$

$$-\frac{(a_{i+1} - a_{i})}{v_{i}\sqrt{(a_{i+1} - a_{i})^{2} + (H/n)^{2}}} + 
\frac{(a_{i} - a_{i-1})}{v_{i-1}\sqrt{(a_{i} - a_{i-1})^{2} + (H/n)^{2}}} = $$

$$-\frac{\cos\varphi_{i}}{v_{i}} + \frac{\cos\varphi_{i-1}}{v_{i-1}} = 0.\eqno(30)$$
Or,

$$\frac{\cos\varphi_{i}}{v_{i}} = \frac{\cos\varphi_{i-1}}{v_{i-1}} = const = \frac{1}{k}.\eqno(31)$$
Or, in the continual limit,

$$\frac{\cos\varphi}{v(y)} = \frac{1}{k} \eqno(32)$$  

However, for $y = y(x)$ it is $\tan\varphi = y'$ and $\cos\varphi = \frac{1}{\sqrt{1 + y'^{2}}}$

Then we have 

$$ x - x_{0} =  \int\frac{v(y)dy}{\sqrt{k^{2} - v^{2}}}.\eqno(33)$$  
We get for $v = Ay$: 
 
$$ x  =  \int_{y_{0}}^{y}\frac{Aydy}{\sqrt{k^{2} - A^{2}y^{2}}} + C\eqno(34)$$ with the solution 

$$(x - C)^{2} + y^{2} = \left(\frac{k}{A}\right)^{2} = R^{2},
\eqno(35)$$  
which is the circle with the centre on the x-axis.

For $v= A/y$, we get instead of equation (34):

$$ x  =  \int\frac{Ady}{\sqrt{k^{2}y^{2} - A^{2}}} + C.\eqno(36)$$ 

Let us remark that the physical meaning of the relation (36) with $v = A/y$ is the problem of the catenary in the homogeneous gravitational field.   

\section{The Poincar\'e optical model of the Loba\v cevskii geometry}

The Loba\v cevskii geometry is the integral part of the general geometry called non-euclidean geometry. The name non-Euclidean was used by Gauss to describe a 
system of geometry which differs from Euclid's in its properties 
of parallelism. Such a system was developed independently by 
Bolyai in Hungary and Loba\v cevskii in Russia, about 120 years 
ago. Another system, differing more radically from Euclid's, 
was suggested later by Riemann in Germany and Schlafli in 
Switzerland. The subject was unified in 1871 by Klein, who gave 
the names parabolic, hyperbolic, and elliptic to the respective 
systems of Euclid, Bolyai-Loba\v cevskii, and Riemann-Shlafli (Coxeter, 1998). 

The substantial mathematical object in the Loba\v cevskii geometry is the  angle of parallelism defined by Loba\v cevskii as follows.
Given a point $P$ and a line $q$. The Intersection of the perpendicular through $P$ let be $Q$ and $PQ = x$. The intersection of line p passing through $P$, with $q$, let be $R$ and $QR = k$. Then, the angle  $RPQ$ for perpendicular distance $x$ 

$$\Pi(x)=2\tan^{-1}e^{-x/k}.\eqno(37)$$
is known as the Loba\v cevskii formula for the angle of parallelism (Coxeter, 1998; Loba\v cevskii, 1914). 

 The Poincar\' e  model of the Loba\v cevskii geometry is the physical model  of the optical trajectories in a medium with the velocity of light $v = Ay$.

 According to Hilbert (Hilbert, 1903), it is not possible to realize the Loba\v cevskii geometry globally on surface  with the constant negative curvature. 
The Beltrami realization of the Loba\v cevskii geometry is only partial.  
 The Loba\v cevskii geometry is the partial  geometry on the pseudosphere with the parametric equations

$$x = a\sin u \cos v, \quad y = a\sin u \sin v, \quad z = a\left(\ln \tan\frac{u}{2} + \cos u\right).\eqno(38) $$
(Kagan, 1947, ibid. 1948; Efimov, 2004; Norden, 1956; Klein, 2004;  Manning, 1963).

 The pseudosphere is the surface generated by the rotation of the Leibniz tractrix with equation

$$x = a\sin u,\quad y = 0, \quad z = a\left(\ln \tan\frac{u}{2} + \cos u\right).\eqno(39) $$

The pseudosphere is in the half geodetic coordinates given by the squared element

$$ds^{2} =  du^{2} + \cosh^{2}\frac{u}{a}dv^{2}.\eqno(40)$$  

 The pseudosphere is in the izothermical coordinates given by the Poincar\' e squared element

$$ds^{2} =  \frac{a(dx^{2} + dy^{2})}{y^{2}}.\eqno(41)$$  

The Leibniz solution of the tractrix problem is as follows:

$$y = a\frac{\ln (a + \sqrt{a^{2} - x^{2}})}{x} - \sqrt{a^{2} - x^{2}}.\eqno(42)$$ 

The trajectory of light in the Poincar\' e model is a trajectory passing from $A(x_{1}, y_{1})$ to $B(x_{2}, y_{2})$  and determined by the minimal time from $A(x_{1}, y_{1})$ to $B(x_{2}, y_{2})$. It  is the result of the minimum of the functional (20, 22).
The Poincar\'e circles in his model are analogue of the for straight lines in the Euclidean geometry. 
The following theorems are valid in the Poincar\' e model of the Loba\v cevskii geometry:

{\bf Theorem 1:}.  Only one half-circle passes through two points $A, B$ in the Poincar\' e plane.

{\bf Theorem 2:} The curvilinear segment $AB$ in the Poincar\' e plane is of the shortest length.
 
{\bf Theorem 3:} The parallels are two half-circles with the intersections on the x-axis.

{\bf Theorem 4:} If point $A \notin q$ then there are $q_{1} \parallel q$, $q_{2} \parallel q$ passing through $A$, with $q_{1} \neq q_{2}$.

{\bf Theorem 5:} If point $A \notin q$, $q_{1} \parallel q, q_{2} \parallel q$, then $q_{1}, q_{2}$ divide the Poincar\' e plane in four different sectors I, II, II, IV.

Let us remark that he optical distance between point $A$ and $B$ is not
equivalent to the mechanical distance realized by the nonelastic flexible fibre as the shortest distance between point $A$ and $B$.
 The Poincar\' e model of geometry where the light velocity is $v  = Ay$ is the interaction model of light with the optical medium. 

It is elementary to see that if we define the Poincar\' e problem on a sphere, then we get so called spherical Poincar\' e model of the Loba\v evskii geometry.
  
Let us use the index of refraction $n(r)$ in the Euclidean plane and polar coordinates $r, \varphi$ to derive the Poincar\' e model.

The the explicit form of the Fermat principle

$$\delta\int n(r)ds = 0\eqno(43)$$  
is (Marklund et al., 2002) 

$$\delta\int n(r)\sqrt{1 + r^{2}\left(\frac{d\varphi}{dr}\right)^{2}} = 0\eqno(44)$$ 

The last equation is equivalent to the Euler-Lagrange variational equation for the functional $F(\varphi, \varphi')$

$$F_{\varphi} - \frac{d}{dr}F_{\varphi'} = 0.\eqno(45)$$ 
Or, 

$$\frac{d}{dr}\left[n(r)\frac{r^{2}d\varphi dr}{\sqrt{1 + r^{2}\left(\frac{d\varphi}{dr}\right)^{2}}}\right] = 0;\eqno(46)$$ 

It is evident that the elimination of $d\varphi/dr$ is as follows:

$$\frac{d\varphi}{dr} = \pm \frac{C}{\sqrt{r^{4}n^{2}(r) - C^{2}r^{2}}} = 0.\eqno(47)$$ 

The circular trajectory is defined by equation $dr/d\varphi = 0$ from which follows the index of refraction as 

$$n(r) = \frac{const}{r}.\eqno(48)$$ 

The experiments has been performed in Bose-Einstein condensate with the result that the optical light pulses can travel with extremely small group velocity about 17 meters per second (Hau et all. 1999). 

\section{Discussion}
 
We explained the Fermat principle in the geometrical optics and in the variational calculus. We  determined  the optical trajectories  in the optical media. We also derived the Poincar\' e model of the Loba\v cevskii geometry. 
The Poincar\' e model of the Loba\v cevskii geometry is the optical model based on the index of refraction, which is the consequence of the interaction of light with medium.
Also, space, when considered independently of
measuring instruments, has neither metric nor projective
properties; it has only topological properties. It is amorphous (Poincar\' e, 1963). That which cannot be measured cannot be an object of science. 
We applied the Fermat principle to the determination of the index of refraction of the medium to get the circular optical trajectories. This effect is the confinement of light by optical medium.

There is no doubt that the monochromatic optical beam is composed
from photons of energy $E = \hbar\omega$. While the rest mass of photon is zero, the relativistic mass follows from the Einstein relation $E = mc^{2}$. After identifying the relativity energy and quantum energy of photon we have

$$m = \frac{\hbar\omega}{c^{2}}. \eqno(49)$$ 

The centrifugal force acting on photon moving with velocity $v$ in optical medium along the circle with radius $R$ is for the photon mass as follows:

$$F_{centrifugal} = \frac{\hbar\omega}{c^{2}} \frac{v^{2}}{R}.\eqno(50)$$ 

The centrifugal force is the origin of the unstability of the photon trajectory in the optical medium. The further origin of the unstability of the trajectory are the thermal fluctuations of the index of refraction. We know, that the Kapitza effect is based on the thermal fluctuations of the index of refraction (Landau, et al., 1982). So, the experimental investigation of the confinement of photon in the optical medium is meaningful at $T \approx 0$. There is no doubt that the investigation of the photon trajectories is the crucial problem of the optical physics and it is interesting for all optical laboratories over the world.

\vspace{9mm}
 
\noindent
{\bf References}

\vspace{9mm}

\noindent
Coxeter, H.S. M.  {\it Non-Euclidean Geometry}, 
The mathematical association of America, Washington,
 D.C. 20036 , 6-th ed., Printed in USA, (1998).\\[2mm]
Efimov, N. V. {\it The Higher geometry}, 7-th ed. Nauka, 
Moscow, (2004). (in Russsian). \\[2mm]
Hau, L. V., Harris, E., Dutton, Z. and Behroozi, C. H, 
Light speed reduction to 17 metres per second in an
ultracold atomic gas, Nature {\bf 397}, 594-598 
(18 February 1999). \\[2mm]
Hilbert, D. {\it The foundation of geometry}, Chicago, London agents, (1902).\\[2mm]
Hilbert, D. (1903). {\it {\" U}ber Fl\"achen von constanter Gauss'schen Kr\"ummung} ({\it On the surfaces of the constant
Gauss curvature}), Giornale di Matematiche, {\bf 6}, p. 1868. \\[2mm]
Kagan, V. F. {\it Foundation of the theory of sourfaces} I,
OGIZ, GITL, Moscow, St. Petersburg, (1947); ibid. II, (1948).
(in Russian). \\[2mm]
Klein, F., Vorlesungen \"uber h\"ohere  Geometrie, 
Dritte Auflage bearbeitet und herausgegeben 
von W. Blaschke, (Moscow), (2004). (in Russsian). \\[2mm]
Landau, L. D. and  Lifshitz, E. M. {\it The classical theory 
of fields}, 7-th ed., Moscow, Nauka, (1988). (in Russian). 
\\[2mm]
Landau, L. D. and  Lifshitz, E. M. {\it Electrodynamics of continuous media}. —2nd ed. —(Course of theoretical physics; 
V. 8), Moscow, Nauka, (1982). \\[2mm]
Lavrentjev, M. A. and Lyusternik, L. A. {\it  Lectures on the variational calculus}, GITL, Moscow, (1950). (in Russian).
\\[2mm]
Loba\v cevskii, N.  {\it The theory of parallels}, La Salle, Illinois, Open court publishing Company, (1914).\\[2mm]
Manning, H. P. {\it Introductory Non-Euclidean Geometry}.
New York: Dover, (1963).\\[2mm]
Marklund, M., Anderson,  D., Cattani, F., Lisak, M.  and 
Lundgren, L. Fermat's principle and variational analysis 
of an optical model for light propagation exhibiting 
a critical radius, ArXiv: physics/0102019v2, 12 Aug 2002.
\\[2mm]
Norden, A. P. Editor, {\it On the foundations of geometries}, Moscow, (1956). (in Russian). \\[2mm]
Poincar\' e, H.  {\it Mathematics and Science: Last Essays} (Dernieres Pensees) Dover Publications, Inc., New York, 
(1963). 

\end{document}